\documentclass[preprint2]{aastex701}
\usepackage{graphicx}

\usepackage{ulem}
\usepackage{color,soul}

\begin{document}

\onecolumngrid
\title{IFU Observations of Comet 3I/ATLAS = C/2025~N1 (ATLAS) using the Hobby-Eberly Telescope with the VIRUS Instrument}  
\twocolumngrid

\footnote{ Based on observations obtained with the Hobby-Eberly Telescope (HET), which is a joint project of the University of Texas at Austin, the Pennsylvania State University, Ludwig-Maximillians-Universitaet Muenchen, and Georg-August Universitaet Goettingen. The HET is named in honor of its principal benefactors, William P. Hobby and Robert E. Eberly. }

\shorttitle{Observations of 3I (ATLAS) with HET and VIRUS}
\submitjournal{PSJ}

\author[orcid=0000-0003-4828-7787]{Anita L. Cochran}
\affiliation{McDonald Observatory, University of Texas at Austin}
\email{anita@astro.as.utexs.edu}

\author[orcid=0000-0003-2307-0629,gname=Gregory,sname=Zeimann]{Gregory Zeimann}
\affiliation{Hobby Eberly Telescope, University of Texas at Austin}
\email{gregz@astro.as.utexas.edu}

\author[0000-0001-8923-488X]{Emmanuel Jehin}
\affiliation{STAR Institute, University of Li\`ege}
\email{ejehin@uliege.be}

\author[orcid=0000-0003-2011-9159]{Hideyo Kawakita}
\affiliation{Koyama Space Science Institute, Kyoto Sangyo University}
\email{kawakthd@cc.kyoto-su.ac.jp}

\author[0000-0002-0622-2400]{Adam J. McKay}
\email{mckayaj@appstate.edu}
\affil{Department of Physics and Astronomy, Appalachian State University}

\correspondingauthor{Anita Cochran}

\begin{abstract}

We observed interstellar object 3I/ATLAS = C/2025~N1 (ATLAS) on 9 December 2025 UT with the VIRUS spectrograph on the Hobby-Eberly Telescope.  VIRUS is a low spectral resolving power, multi-IFU, optical spectrograph. We used a single IFU for the observations and were able to obtain spectra spaced every one arcsec (1327\,km) from the optocenter out to about 85,000 km in the sunward direction. The spectra include molecular features normally seen in 
the optical for spectra of Solar System comets, with the addition of many very strong lines of FeI and NiI.  We modeled these observations using the Haser (1957) model framework and
find that the data can only be fit if we assume that the outflow velocity of the gas is half that of comets in our Solar System 
observed at the
same heliocentric distance. We derived Haser model scale lengths for three NiI and four FeI lines. We discuss the implication of the similarities and differences between this object and normal Solar System comets.
The excellent spatial resolution of VIRUS, coupled with a FOV of $50\times50$\,arcsecs and a geocentric distance of 1.83\,au, allowed us to probe the coma of 3I in great detail. 
An accidental offset of the optocenter of the comet to the edge of the FOV meant that we probed to a large distance ($>$85,000km) in the sunward direction from the optocenter, adding new information to the extensive literature on 3I, even with a single epoch of observations.  
\end{abstract}

\section{Introduction} 

We study comets in our Solar System as windows on conditions in the early stages of planet formation.  Comets are the least altered bodies that remain from that time \citep{MummaCharnley2011ARAA,altweggetalaraa}.  Our current understanding is that some comet precursors formed in the region beyond Neptune; those leftovers include the population of Jupiter Family Comets (low inclination objects with orbital periods of less than 20 years) and the scattered disk population \citep{Donesetal_cometbook_2004, Nesvorny2018ARAA}.  Other comet precursors were also formed in the giant planet region.   Some of these were incorporated into the growing giant planets but, for those not included in building the giant planets,  5--10\% were gravitationally perturbed into the Oort Cloud and 90--95\% of them were ejected completely from the nascent Solar System \citep{FernandezBrunini2000,Kaib2024cometbook}.

It is logical to assume that other planetary systems went through similar formation and sorting processes, leaving some small objects in these other planetary systems but ejecting other ``comets" out into the Galaxy \citep{Morbi_2008}.  Some of those ejected comets are on trajectories crossing the Solar System.  Such extrasolar comets can be identified by hyperbolic orbits and high excess velocity (\(v_{\infty }\)).

At present, there are three small objects that have been discovered and classified as interstellar objects.  
The first such object, C/2017 U1 = 
1I/'Oumuamua, was discovered with the Pan-STARRS telescope by Robert Weryk on 2017 Oct. 19 \citep{Meech_1I}. The nucleus was quite small and elongated.  It did not have any obvious gas emissions or coma, but it did exhibit non-gravitational motions to its orbit \citep{Michelietal2018}.

The second interstellar body was discovered on 2019 Aug. 30 by amateur astronomer Gennadiy Borisov and is designated C/2019 Q4 = 2I/Borisov.  Borisov showed a gaseous coma whose composition exhibited high CO relative to H$_2$O compared with Solar System comets \citep{jewittLuu2I}.

The last discovery,  on 1 Jul. 2025, was 3I/ATLAS = C/2025~N1 (ATLAS) (hereafter, we will designate this object as 3I). By studying 3I, we hope to learn about conditions in the natal planetary system of this target and/or its long interstellar journey.

Observations of interstellar objects are important to our understanding of the formation of our Solar System.  This is true regardless of whether they look compositionally just like typical Solar System comets, or look very different, because it speaks to the universality of the processes by which comets are formed.  However, it must be remembered that we do not know the conditions under which the interstellar objects we have discovered were formed, how old the objects are, what type of stellar system they formed with and how the
structure of the nucleus has been
altered as the object wanders around the Galaxy.
Indeed, \citet{hopkinsetal2025} concluded that 3I must be at least 7.6\,Gyr, based on the velocity and radiant of its orbit.  They argued for a different origin for 3I than for either of the previous two interstellar objects.
Based on kinematic calculations, \citet{TaSe2025} figured that the age is somewhere between 3 and 11\,Gyr.   From isotopic considerations, \citet{2026arXiv260306911C} figured that 3I was formed 10-12\,Gyr ago.  Thus, while it is quite possible that 3I is older than our Solar System, its age is not definitively known.

\citet{2026arXiv260306911C} and \citet{Opitometal2026} examined the isotopic ratios for carbon and nitrogen in 3I and compared them with bodies in our Solar System.   $^{14}$N/$^{15}$N is two times higher in 3I than in Solar System comets, while $^{12}$C/$^{13}$C is moderately higher.
D/H in 3I is substantially higher than normal comets, with the D/H in H$_2$O and CH$_4$ being measured as 10--40 times higher than Earth or comets
\citep{2026arXiv260320445R,2026arXiv260307026S,2026arXiv260306911C}

Observations of 3I show that it is very enriched in CO and CO$_2$, with a CO$_2$/H$_2$O mixing ratio higher than previously measured comets \citep{2025ApJ...991L..43C, Lisse_spherex1, Lisse_spherex2}. 
\citet{2025ApJ...991L..43C} concluded that 3I is
either intrinsically CO$_2$ rich or is H$_2$O ice depleted.  This has implications for the formation temperature of the body.

\citet{maggioloetal2026} concluded that substantial galactic cosmic rays (GCRs) impinged on its surface, altering the structure. They argued that these objects have heavily processed surfaces, rather than pristine ones. Whether by GCRs or by differences in insolation, the structure of the surface ices might be quite different from bodies in our Solar System, accounting for observed spectral differences
such as those presented in this paper.

\section{Observations}
We used the Hobby-Eberly Telescope and its wide-field, low resolving power spectrograph, VIRUS \citep{HillVirus2021}, to obtain a set of data on 3I.{\bf \footnote{The Hobby-Eberly telescope (or HET) is a 10m telescope sited on Mt. Fowlkes as part of the McDonald Observatory complex outside of Ft. Davis, TX.}}
The data were obtained on 2025 December 9 UT. At the time of the observations, 3I was at a heliocentric distance of 2.03\,au and a geocentric
distance of 1.83\,au and was moving away
from the Sun.

VIRUS stands for Visible Integral-Field Replicable Unit Spectrograph, so named because it contains 78 pairs of spectrographs that are
identical in order to save costs and gain field of view.
Together, the IFUs cover a field of view (FOV) that is 18 arcmin on a side. VIRUS spectrographs cover 3500--5500\AA, with R = $\lambda/\Delta \lambda = 700$. 
The fibers are 1.5 arcsec in diameter and each IFU covers 50 arcsec by 50 arcsec on the sky.  There are spaces between
adjacent fibers. Those spaces are filled, and a field is fully imaged, using 3 separate pointings (dithers; \citet{HillVirus2021},
see their figures 7 and 8).

The HET is an unusual telescope \citep{RamseyHET} because it is fixed in altitude. Its azimuth can be changed, but once at the desired azimuth, the telescope structure does not move during observations.
Instead, a tracker at the prime focus is used to trace the path of the target across the sky. Two guide cameras offset from the VIRUS FOV can lock onto stars to ensure that the telescope stays pointed at the target.  However, for objects moving at non-sidereal rates, the guide stars do not stay fixed with respect to the prime target. Thus, the non-sidereal motion must be compensated for, and  the guide star(s) must be handed off from one guide camera to the other.  

There was some interval of time between when the target was sent to the IFU and the guide stars were set up and ready to follow 3I with appropriate tracking.  As it turned out, the interval for the hand-off was not properly accounted for, so 3I ended up
on the edge, and not the center, of an IFU.  However, once on its guiding location,
the telescope tracked the moving target well, venturing off tracking 3I correctly by 1 arcsec or less over the three dither observations (just under 20 minutes).  The errors in tracking were checked by comparing the three
dither sub-frames.  

Since the comet optocenter ended up on the edge of the IFU, we ran several tests to confirm that the optocenter was on the detector and we knew on which pixel it was imaged.  One test was to fit contour plots that showed the peak was near, but not off, the edge of the detector. Another test was to derive column densities assuming different optocenter pixels and determine if the data were symmetric around the selected pixel and which pixel had the maximum counts for all features.  We were able to convince ourselves that we knew the location of the optocenter to within 1 pixel and the optocenter was just on the IFU, but at the edge.

The data reduction was performed with the package ``Remedy" (see \citet{ZeimannVIPS} for complete details on the data reduction). 
Basically, typical IFU handling procedures are followed including bias and dark removal, removal of bad pixels, gain correction, sky removal and extraction and appropriate combinations of fibers. All of the data are formed into cubes, accounting for differential atmospheric refraction, 
with X and Y on the sky and wavelength defining the slices. 

The wavelength scale for each spaxel was defined with observations of a HgCd lamp. All of the spectra were wavelength rectified onto a common linear grid of 3470 -- 5540\AA\ with 2\AA\ per slice.  There are
1036 slices (wavelength intervals) per output target cube. 

Flux calibration was performed using a fixed empirical instrument response curve derived from spectrophotometric standard-star observations obtained over an extended period. For each exposure, this response curve is scaled using the measured mirror illumination and guider throughput to account for variations in telescope throughput and observing conditions. This procedure provides an absolute flux calibration that is typically accurate to about 15\%. The relative flux calibration, referring to the accuracy of the spectral shape across wavelength rather than the overall normalization, is more precise, with an accuracy of approximately 5\% above 3900\,\AA\ and 10\% below 3900\,\AA. During cube construction, fiber flux densities are divided by the fiber area and resampled onto a regular spatial grid using a flux-conserving reconstruction. Because the pixel area is explicitly accounted for, the resulting values correspond to flux density per square arcsec for a $1'' \times 1''$ grid.

The solar spectrum was constructed from the HETDEX Continuum Catalog (version 5.0.1; \citet{Mentuch_Cooper_2023}). We selected G-type stars and computed a biweight average spectrum on a fixed 3470–5540\,\AA\ grid, normalizing by the g-band AB flux.

\begin{figure}[h]
\includegraphics[width=0.45\textwidth]{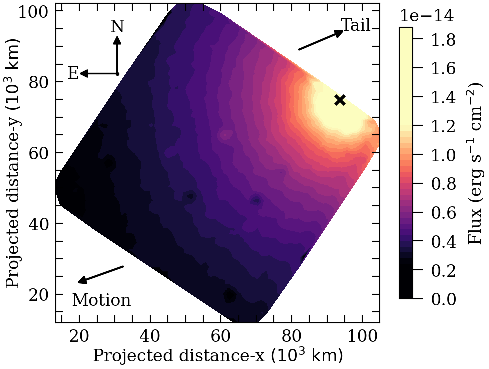}
    \caption{The spatial distribution of the CN fluxes on the IFU FOV is shown. The position of the optocenter of 3I, on the edge of the IFU, is denoted by an X.  The CN fluxes are determined every 1 arcsec, or 1327km.   The tail direction marked is just the extended heliocentric radius vector as we did not actually see a tail.  We mark the direction of motion. \label{image}}
\end{figure}

Figure~\ref{image} shows the spatial distribution of the summed fluxes in the CN bandpass (3865-3905\AA) for the three combined dithers. The comet optocenter is on the right edge of the IFU. Although the placement on the edge of the array
was unfortunate, the reasonably accurate tracking meant that we could still use these spectra to determine what gases were present in the spectra and how they behaved with cometocentric distance.

\begin{figure*}[h]
  \centering
\includegraphics[width=0.75\textwidth]{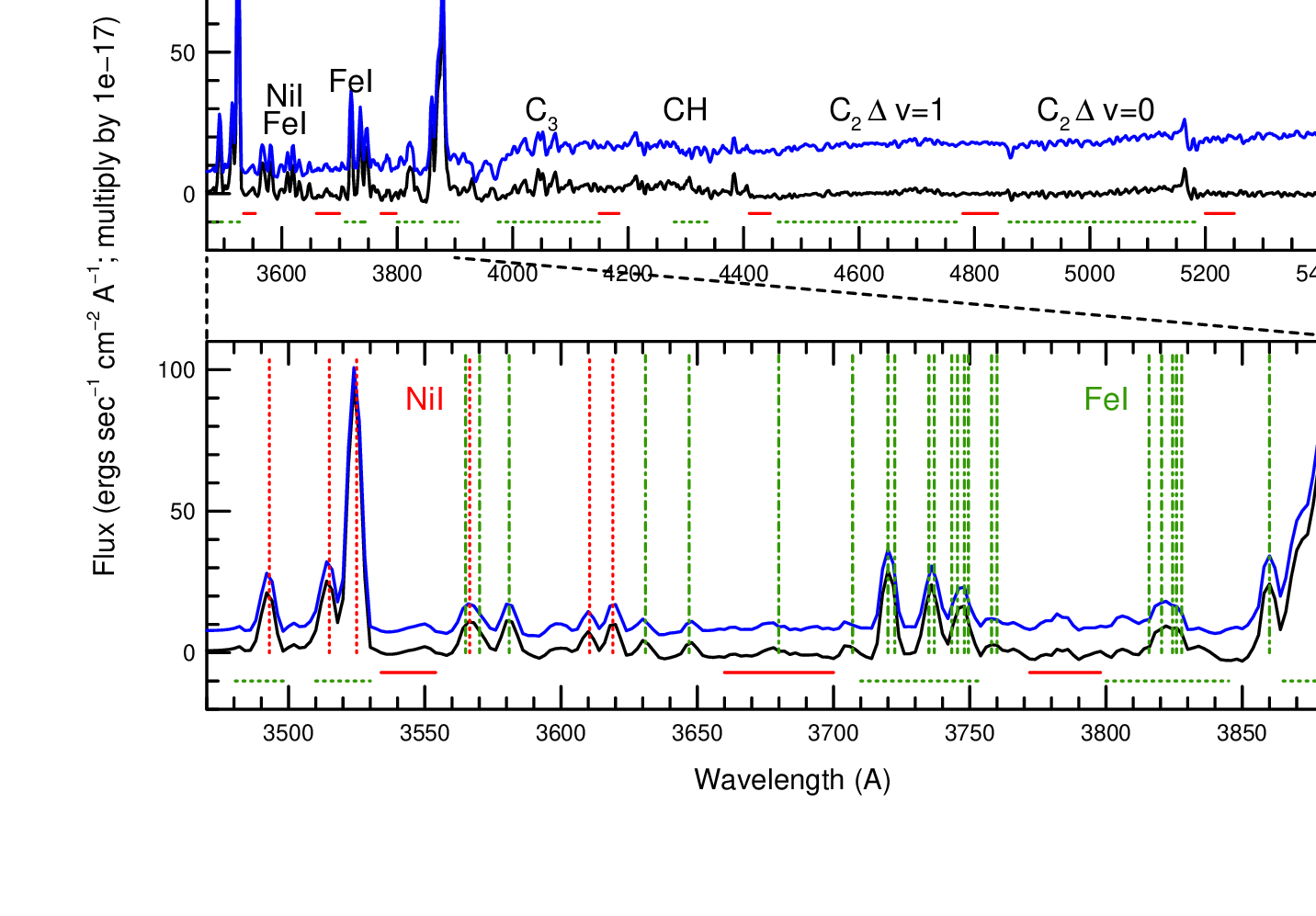}
\caption{A spectrum from a spaxel 
near the optocenter is shown.  The complete spectrum is displayed in the upper panel, with features identified.  The blue spectrum is the comet spectrum with the solar continuum spectrum reflected off cometary dust imposed on the comet spectrum.  The black spectrum is the
same spectrum but with the dust removed (see section 3). The lower panel is an expanded view of the bluer end of the spectrum.   There are many atomic NiI and FeI features identified in the lower panel with the NiI lines as red dotted lines and the strongest FeI lines as green dashed-dotted lines.
On both panels, the horizontal green  dotted lines under the spectra delineate the feature bandpasses used, while the horizontal solid red lines under the spectra delineate the continuum regions used.
\label{spec}}
\end{figure*}

Only one IFU out of the VIRUS array of 78 pairs of spectrographs contained an appreciable signal from 3I.   However, that one IFU imaged a $50\times50$\,arcsec
patch of the sky with thousands of data points,
making these data a very rich, and unique, look at the gas in the inner $\sim85,000$\,km
of the coma.
Figure~\ref{spec} shows the observed spectrum from 1 spaxel near the optocenter. Various molecular and atomic features are identified. 
The region of the spectrum redward of
3850\AA\  shows the usual molecules at the
same relative strengths that we normally detect in the spectra of
comets observed in the inner Solar System.   In 3I, we detected CN, C$_3$, CH and two bands of C$_2$.  These species result from the dissociation of various ices that are in the nucleus, as well as chemistry in the inner coma.

\section{The spatial distribution of gas in the coma}
At the end of the data reduction, we have spectra at every wavelength and spaxel. We want to determine what the gas is doing as it flows away from the nucleus, so we need to convert these spectra into useful information at each position.  

The spectra after reduction are a mixture of cometary emission features and a continuum generated by solar photons reflecting off the dust.  Since we wish to measure the gas distribution, we need to remove the solar/continuum spectrum first.  We defined seven continuum regions, marked underneath the spectra in Figure~\ref{spec}, and computed the sum of the flux in each of the continuum regions in both the comet spectra and a solar spectrum.  The ratio of those two sums then yields a factor which we
used to scale the solar spectrum to the continuum of the comet spectrum.  With only seven continuum regions, we needed to fit a cubic polynomial to these ratios in order to define a factor at each wavelength to scale the solar spectrum to the comet.  Then we just subtracted the scaled solar spectrum from the comet plus dust spectrum.  That yielded a spectrum that contained only cometary gas features.
It was these comet gas spectra that we used to determine the quantity of each gas species at each position on the detector.

\subsection{Molecules}
In order to place the data on a scale that could be compared with Solar System comets, we need to convert from the flux at each wavelength to the column density for a particular gas species.  First, we sum the flux within bandpasses that encompass the whole band
for a given molecule.  Then we convert this integrated flux to a column density using standard efficiency factors. 
We used the efficiency factor values listed in Table~1 of \citet{coetal2012}.  

\begin{figure}
\centering
\includegraphics[width=0.45\textwidth]{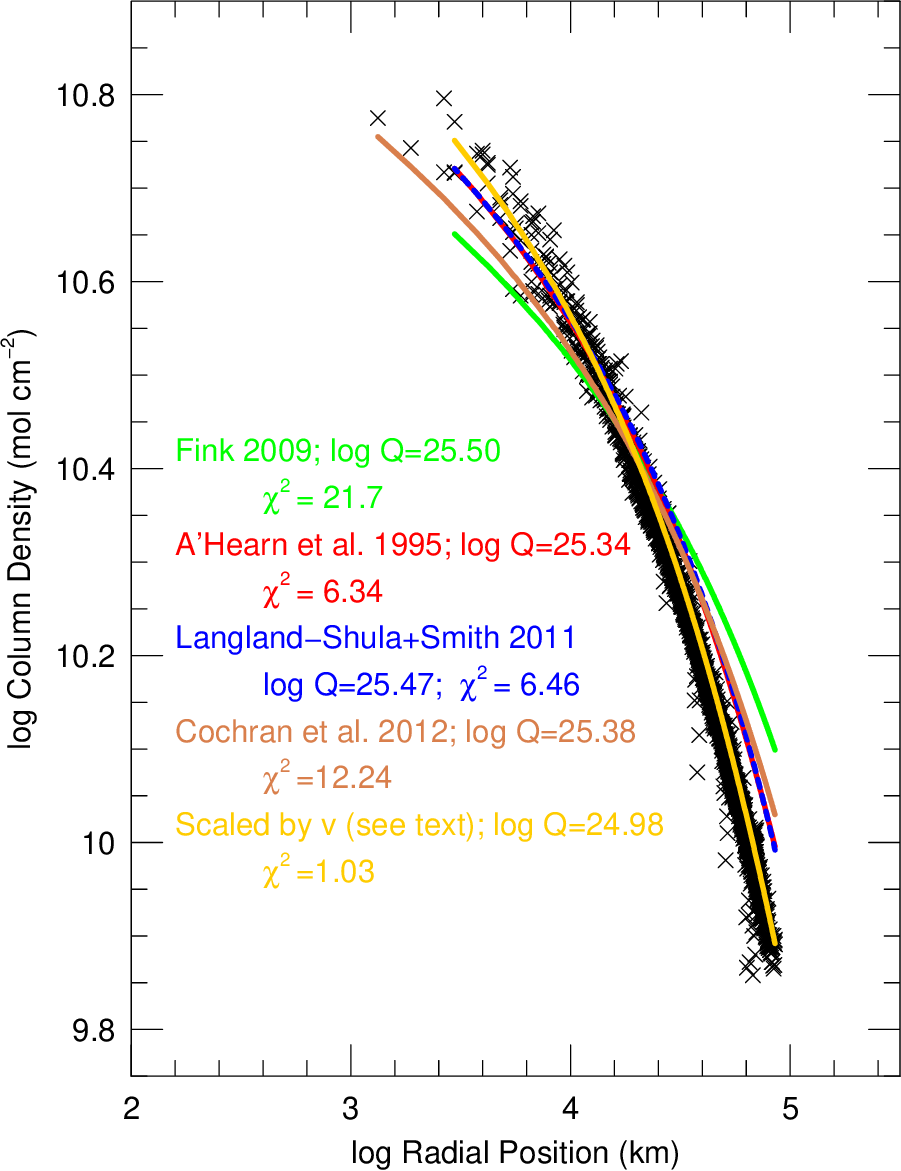}
    \caption{The column densities for CN  are shown as a function of distance to the optocenter.  We computed Haser model production rates (see text) for each molecule's distribution.  As discussed in the text, there are several widely used scale lengths to convert from column densities to production rate. Fits for several  scale lengths from the literature are shown, along with a set of scale lengths that were scaled from a standard set (see text).\label{CNscale}}
\end{figure}

The distribution of
the column densities for the CN molecule is shown in Figure~\ref{CNscale}.
Inspection of this figure shows that we have abundant column density points to define the distribution of CN.  With the comet optocenter at the edge of the FOV, we preferentially were looking on the sunward side of the nucleus.  The placement of the optocenter on the edge of the field meant that our observations measured intensities at large distances from the nucleus. For all of the common molecular species, we traced the gas out to 
$\sim$85,000\,km.  The profiles include points offset at different angles from the nucleus. We see no real difference in the gas distribution with angle. 

To convert the observed column densities to production rates for comparison with Solar System comets, we used the Haser model \citep{ha57}.  This approach is widely used in the comet literature and provides a common framework for reporting production rates 
(for example: \citet{ahetal95, fi09, lasm11, coetal2012}). 
The model assumes a steady, spherically symmetric coma with gas expanding radially outward at a constant velocity and with photodissociation scale lengths that are fixed for the observing geometry. Our data show that the coma morphology is reasonably symmetric, making these assumptions a pragmatic approximation for the present analysis. From an inference standpoint, the Haser model could be viewed as a simplified forward model that encodes a particular set of assumptions about the structure of the coma and the photochemistry. However, these assumptions are not expected to hold exactly for real comae, which can exhibit anisotropic outflow, time variability, and more complex chemical pathways. For this reason, the derived production rates should be interpreted cautiously as model-dependent quantities, rather than direct physical measurements.

The Haser model can be expressed as
\begin{equation}
    n = Q[4\pi r^2v]^{-1} [\beta_0/(\beta_1 - \beta_0)] [e^{-\beta_0r}-e^{-\beta_1r}]
    \label{haser}
\end{equation}
where n is the number density of the daughter species, $\beta_0$ and $\beta_1$ are the reciprocal scale lengths of the parent and daughter species respectively, $v$ is the outflow velocity, Q is the production rate of the parent and r is the radial position within the coma.
The scale lengths for the Haser model are generally reported in the literature at
1\,au.  When running the models discussed here, 
the scale lengths were scaled to the heliocentric distance of the comet observations
using a factor of R$_h^2$ (where R$_h$ is the comet's heliocentric distance in au) for most molecules and scale lengths in the literature.  The exception is that the C$_2$ parent scale lengths are scaled by R$_h^{2.5}$ for the scale lengths of \citet{coetal2012}.
The outflow velocity used for our modeling was $v = 0.85\,R_h^{-0.5}$ \citep{cosc93}.
At the heliocentric distance of 3I, this relation yields $v$=0.597\,km/s.   Inspection of equation~\ref{haser} shows that the scale lengths define the shape of the gas distribution with distance to the optocenter.  Q and $v$ control the overall total number densities and are dependent on one another. Thus, changing $v$ changes the derived Q. 

Figure~\ref{CNscale} shows the CN data with model fits from four common sets of scale lengths in the literature \citep{ahetal95, fi09, lasm11, coetal2012}, identified by color (since the \citet{ahetal95} and \citet{lasm11} models completely overlap, their lines are dashed).  The fits are normalized at a log column density of 10.4. The narrowness of the distribution of the CN column density with cometocentric distance is indicative that the error bars on each data point is small. Analysis shows that for CN, the error bar on the log column densities would be
much less than $\pm0.1$ at each point. The error bars are computed from 1-$\sigma$ per-sample errors that are
propagated through the imaging and Gaussian
smoothing.

Inspection of this figure shows that none of the models using published scale lengths are good fits to the data.  
As noted above, the scale lengths control the shape of the distribution. The scale lengths depend on the gas being studied and the target's distance to the Sun. The scale length expresses the distance a molecule will travel in the coma in some unit of time.  Thus, if the velocity of the outflowing gas changes, not only will $v$ change in equation~\ref{haser}, but the scale lengths will also change.  

It was not possible to determine the outflow velocity of the gas from our data owing to the low
spectral resolving power of VIRUS. \citet{biveretal2026} used the IRAM radio telescope and observations of HCN and CH$_3$OH to determine that the outflow velocity was 0.37\,km/s when the comet was at a heliocentric distance of 1.36\,au. The velocity equation above would have predicted $v=0.73$\,km/s at the heliocentric distance of their observations.  Scaling their velocity by the square root of R$_h$ to our observation's distance of 2.03\,au would predict a velocity of 0.31 km/s, not the canonical 0.597\,km/s.

\citet{2026arXiv260306911C} used ALMA observations to measure the velocity when they observed 3I at R$_h = 2.4$ au. They found that the outflow velocity of the CO gas was 0.345\,km/s and of the HCN gas was 0.276\,km/s.  At the heliocentric distance of their observations, we would have expected $v$=0.55\,km/s.  Scaling their velocities to our
heliocentric distance would yield 
$v$=0.38 km/s with the CO velocity value or 0.30 km/s with the HCN velocity. 

In all three cases, the gas was flowing away from the nucleus at about half the expected speed.
If we assume that these slow velocities change the scale lengths of both the parent and daughter species, we can run the Haser model with scale lengths that are adapted to the more
slowly outflowing gas. Since the scale length is a
function of the velocity and the molecular lifeime, if the molecule being studied remains the same as in our typical comets, then so will its lifetime. Thus, the scale lengths will change by the ratio
of the velocities.    With a velocity of 0.30\,km/s (as predicted based on scaling the radio telescope velocities) and assuming the same molecular lifetimes, the scale lengths should change by a factor of $\sim$two.  

We computed the model with this slower velocity and the resultant different scale lengths.  The derived curve is shown in Figure~\ref{CNscale}.  This is a much better fit than any of the models with the published scale lengths.  
If we assume
that only the parent outflow velocity (and therefore the parent scale length) changed by a factor of 2, but that the outflow velocity for the daughter was the normal value for gas flowing away from a comet in our Solar System, we find that the fit is not as good as when we change the outflow velocities for both the parent and daughter.  The reduced $\chi^2$ goes from 1.03 for changing both scale lengths to 1.75 for changing just the parent.

\begin{figure}
\centering
\includegraphics[width=0.45\textwidth]{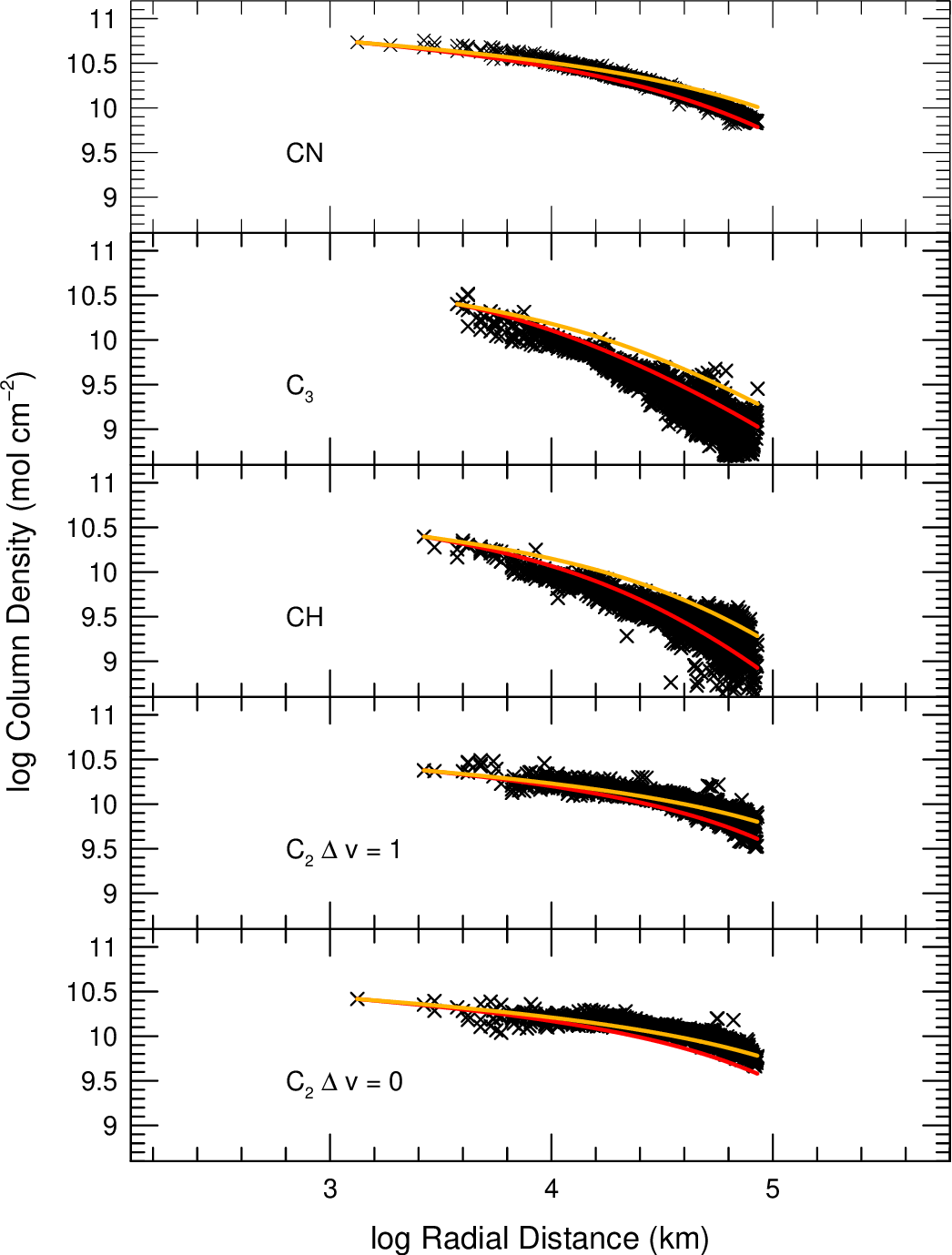}
    \caption{The column densities for CN, C$_3$, CH, and C$_2$ (two bands) are shown as a function of distance to the optocenter.  Models using the scale lengths of \citet{coetal2012} are shown with orange lines.   Models using scale lengths that
    are scaled by the low velocity of 3I (0.3 km/s) are shown with red lines.
    The error bars in the inner coma are smaller than the point markers.  Even the least reliable column densities, those for C$_3$ and CH, have log column density error bars less than $\pm 0.1$ for the data at the largest distances from the optocenter.}
    \label{normal}
\end{figure}

The distribution of gas with distance to the nucleus is shown  in Figure~\ref{normal} for the other species  that are commonly seen in optical spectra of Solar System comets. 
The data are shown with Haser models using the 
canonical scale lengths of \citet{coetal2012}.
The models are shown in orange.

There is no reason to assume that only the CN gas is moving slowly (especially since the slow gas was measured in the radio in several molecular species). Thus, we reran the models with scale lengths that were half of the \citet{coetal2012} ones due to the ratio of normal to slower velocity.
These models are shown in red in Figure~\ref{normal}.  

The reduced $\chi^2$ values for the fits indicate
that C$_2$ modeled with the scale lengths that we  show in Table~\ref{moldata}
are only slightly better than models with the scale lengths of \citet{coetal2012} (less than 10\%).  
On the other hand, the CH and C$_3$ fits with the scaled velocity and, therefore, scale lengths show considerable improvement.  The  CH fit is improved by almost a factor of 2 in $\chi^2$, while the $\chi^2$ value for the C$_3$ fit is 25\% lower for the scaled parameters than for the literature values.  The CN data are less noisy than other species. That makes the fits more sensitive to shape differences.  The scaled fit 
has a reduced $\chi^2$ value 10 times lower than the fit with the normal \citet{coetal2012} parameters,
as can be seen in Figure~\ref{CNscale}.

Table~\ref{moldata} gives the values for the scale lengths we used (at the distance of 3I from the Sun and with the outflow velocity of 0.3\,km/s).  We also show the derived production rates. 
\begin{table}
\centering
\caption{Haser Model Scale Lengths Used\label{moldata}}
\begin{tabular}{lccc}
\hline
 \multicolumn{1}{c}{Molecule} & Parent & Daughter &  log Q \\ 
  & (km) & (km)  & (mol/s) \\
\hline 
CN & 3.5e4 & 6.2e5 &24.98$\pm$0.02 \\ 
C$_3$ & 6.5e3 & 3.0e5 & 24.15$\pm$0.06 \\
CH & 1.6e5 & 1.0e4 & 25.59$\pm$0.07 \\
C$_2 \Delta v=1$ & 7.4e4 &  2.5e5 &  25.04$\pm$0.04 \\
C$_2 \Delta v=0$ & 7.4e4 & 2.5e5  & 25.05$\pm$0.04 \\
\hline \vspace{5pt}
\end{tabular}
\end{table}

Other authors observed some of the same species
\citep{2026arXiv260116983H,
2025arXiv251209020H, 2025arXiv251011779H,
2026arXiv260223586M,
2026arXiv260307718Z,Rahatgaonkaretal2025}
but just used literature scale lengths to derive production rates. Most do not include spatial information in their publications so it is difficult to assess the quality of the fits of their models to their data.  \citet{2025arXiv251011779H,2026arXiv260116983H}
used a similar instrument to us (KCWI on Keck), but their FOV was limited to $<10,000$\,km (log radial distance = 4). They would not be sensitive to deviations of the model from the data with observations only this close to the optocenter.

We can use the production rates measured to compare 3I to Solar System comets.  The production rates that were reported in the cited papers are not entirely consistent with one another, generally because of the use of different scale lengths. There seems to be some consensus that C$_2$/CN looks somewhat depleted relative to ``typical" comets in our Solar System.    A caveat is that one must use the definitions of ``typical" and ``depleted"  that go with the scale lengths that are used for the interpretation of the data. 
\citet{2026arXiv260307718Z} and \citet{2025ATel17538....1J} found that 3I started off depleted in C$_2$/CN but became less depleted (or even not depleted) after perihelion. Our observations were during their post-perihelion observations, when the comet was not depleted in C$_2$.

\citet{coetal2012} defined ``depleted" values for Solar System comets as log(Q(C$_3$)/Q(CN))$<$-0.86 and log(Q(C$_2$)/Q(CN))$<$0.02. 
We found values for these two quantities of
-1.0 (C$_3$) and 0.15 (C$_2$) using the scale lengths of \citet{coetal2012} and -0.83 (C$_3$) and 0.07 (C$_2$) with the scale lengths we derive in this paper.
Our observations do not show the C$_2$/CN being depleted, although it is at the low end of the ``typical" regime. 
Our C$_3$/CN does look somewhat depleted.

Inspection of Figure~\ref{normal} shows that the distributions of the usual molecules with cometocentric distance look fairly symmetric.  With the placement of the comet on the array, many of our data points are more or less sunward.  However, towards the edge that the comet was placed on, those data are along lines that are perpendicular to the Sun-tail line.  These perpendicular vectors cannot be differentiated from the purely solar vector points, confirming that the coma looks symmetric in our data.  However, with the comet's 
placement on the edge of the IFU, we cannot determine if the comet is also symmetric in the tailward direction.

The molecules that we detected are daughter species and not the ices making up the nucleus.  Observations with JWST were used to study the coma formed from parent species.  
\citet{2025ApJ...991L..43C} found that CO$_2$ was enhanced in the sunward direction. 
\citet{2026arXiv260320460R} showed that some species were symmetric and some were not (anisotropic for apolar species and isotropic for polar species; see their Figure 2). 

\subsection{NiI and FeI}
The spectrum blueward of 3850\AA, shown in the lower panel of figure~\ref{spec}, does not look like a typical Solar System comet because we observe obvious features,  identified as NiI and FeI, which are not seen as strong lines in similar spectra of Solar System comets.  Until this century, the only comets for which FeI and NiI were observed were the Great Comet of 1882 and comet C/1965~S1 
(Ikeya-Seki), both sungrazing comets
(see \citet{manetalFeNi2021} and
references therein).
With the advent of better detectors and larger telescopes, \citet{manetalFeNi2021} showed that
FeI and NiI are present in the spectra
of many comets at a wide range of heliocentric distances.  However,
the detected lines are relatively weak and can
only be seen at high spectral resolving power 
(very weak lines that are observable at high spectral resolving power cannot be detected at low spectral resolving power 
since the small line flux gets spread over a wider resolution element and disappears below the continuum). Inspection of a very large number of
low spectral resolving power spectra of comets (those of \citet{coetal2012}) showed no detections of these lines. However, in Figure~\ref{spec}, these lines are very obvious in the spectrum of 3I.  Indeed, the NiI line at 3525\AA\ appears brighter than the CN feature in this spectrum.

Since FeI and NiI were not typically detected at low resolving power, there were no computations of the fluorescence efficiencies needed to compute the column densities until 2021, when \citet{brometal2021} published calculations for many NiI and FeI lines. These calculations included the effect of the heliocentric radial velocity of the comet shifting the emission lines from the comet into and out of alignment with the solar absorption lines of the continuum. When solar absorption features and a cometary emission line are at similar wavelengths, the cometary line sees less solar exciting flux.  

\begin{figure}
    \centering
\includegraphics[width=0.45\textwidth]{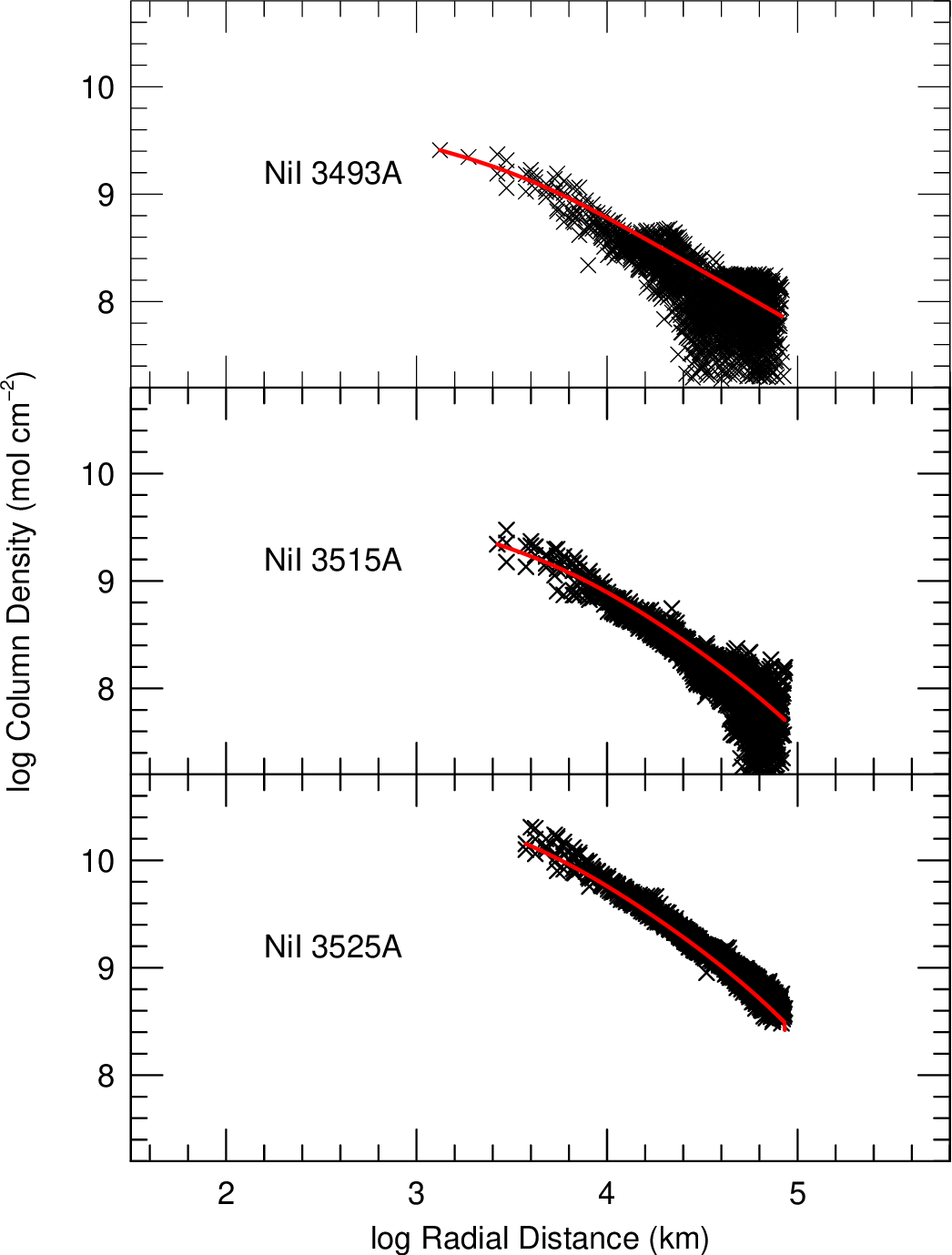}
\caption{The column densities of three of the NiI lines are shown as a function of cometocentric distance.  The model fits are marked with a red line. Again, the error bars for most of these data are smaller than the data points in the plot for the inner coma. The largest error bars are $\pm0.375$ far from the optocenter for Ni\,3493\AA.}  \label{nidata}
\end{figure}

\begin{figure}
    \centering
\includegraphics[width=0.45\textwidth]{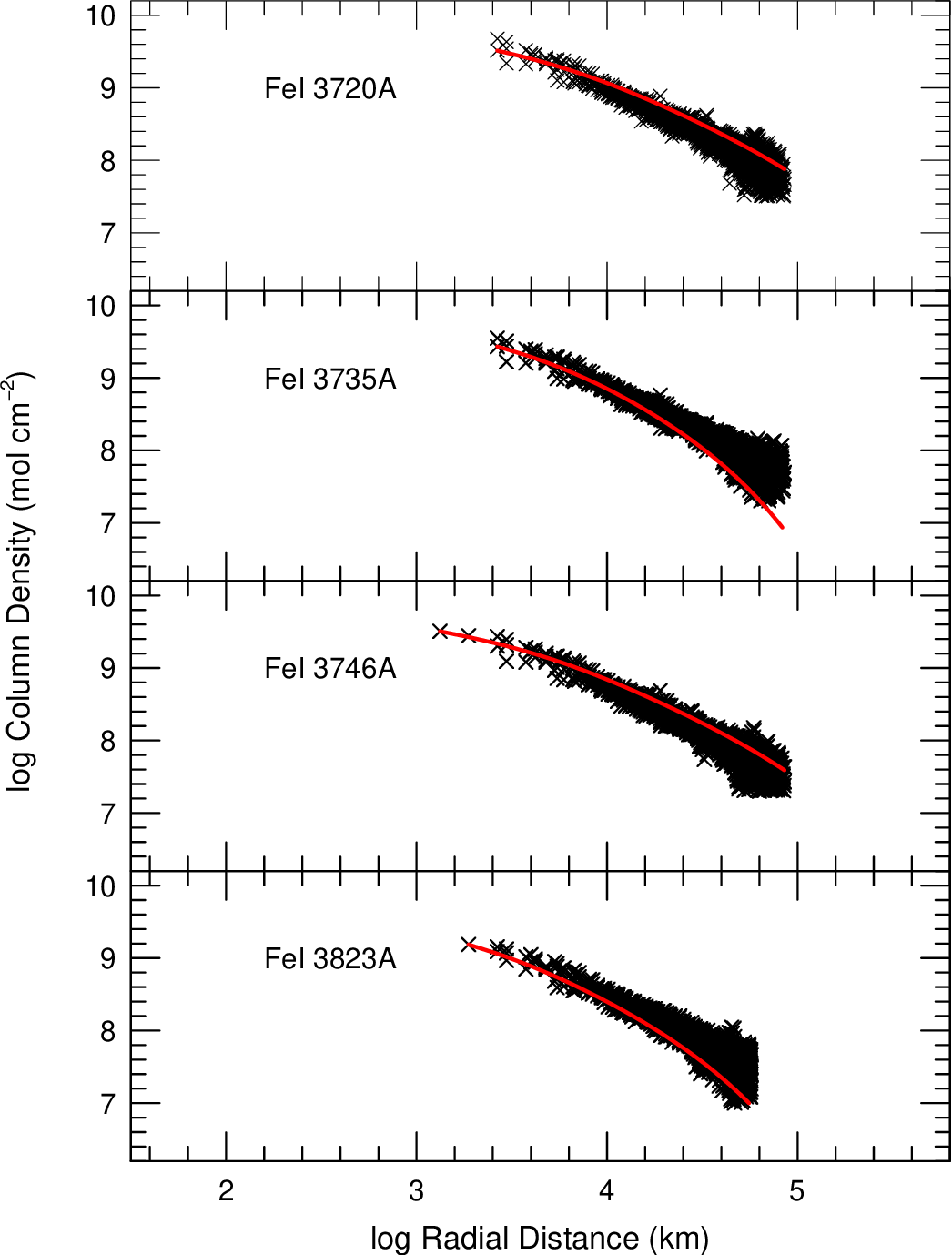}
\caption{The column densities of four  groups of FeI lines are shown as a function of cometocentric distance.  The model fits are marked with a red line. The error bars are again quite small, with the largest error bars being that in the outer regions of the Fe 3746\AA\ line, which is $\pm0.15$} \label{fedata}
\end{figure}

We measured
the strength of each line above the continuum for NiI lines and groups of lines for strong FeI lines to determine the integrated flux
for each line or groups of lines. We then used 
the \citet{brometal2021} fluorescence efficiencies (using the updated version of the code archived on Zenodo \citep{bromley_2026_20514283}; Bromley personal communications)  to convert the fluxes we measured for three NiI lines and four FeI line groups to column densities at each spaxel.  For the FeI line groups, we summed the fluorescence efficiencies of the stronger lines to convert fluxes to column densities.
The column densities are shown in Figures~\ref{nidata} and \ref{fedata}.

Note in these figures that the column densities of individual lines of a species at different wavelengths do not all have the same maximum.  Differences in fluorescence efficiencies result in different strengths for lines of the same molecule (see Figure~\ref{spec}).  In addition,
depending on the heliocentric radial velocity of the comet, some lines will experience different amounts of fluorescence energy from the solar spectrum (the so-called Swings efffect).  Also, for the Fe lines, all of these column densities include multiple 
lines from different transitions (see the notes to Table~\ref{feniscales}; this is a complication caused by our low spectral resolving power).  
To the extent that the fluorescence efficiencies of the different transitions are correct, the different lines might contribute
more or less to the combined line fluxes. Fe lines that might not have been included, or were added at a wrong assumed strength, will make
some of the Fe line column densities less accurate.
However, the spatial distribution of the photons of a particular line will be accurate for that one line and, thus, 
the scale lengths we derive should be considered 
more correct than the overall production rates for Ni and, especially, Fe.


As these species are not seen in low resolving power observations of Solar System comets, there are no published scale lengths for Haser models of these species. We computed grids of models with our Haser code to determine a set of scale lengths for the data in Figures~\ref{nidata} and \ref{fedata}.  Best fits were defined as those with the smallest $\chi^2$ values.
The scale lengths are listed in Table~\ref{feniscales}.
\begin{table*}
   \centering
 \caption{The Derived FeI and NiI Scale Lengths (at 2.03\,au \label{feniscales}}
 \begin{tabular}{ccccccccc}
    \hline
   \multicolumn{1}{c}{Wavelength} & Parent & Daughter & Log Q & 
   \hspace{1in} &
   \multicolumn{1}{c}{Wavelength} & 
   Parent & Daughter & Log Q \\
   \multicolumn{1}{c}{(\AA)} & (km) & (km) &
   (mol/s) & &
   \multicolumn{1}{c}{(\AA)} & (km) & (km)
         & (mol/s)  \\
    \hline
    \multicolumn{4}{c}{\bf FeI} & \multicolumn{4}{c}{\bf NiI} \\
    \hline
    3719.7, 3722.5 & 4.12e3 & 2.06e5 & 23.10$\pm$0.05 &  &
    3493.0 & 2.97e3 & 3.30e7 & 22.76$\pm$0.08 \\
    3734.9, 3737.1 & 2.97e3 & 4.12e4 & 23.41$\pm$0.15 &  &
    3515.1 & 4.12e3 & 2.06e5 & 22.98$\pm$0.06 \\
    3746$^a$ & 3.13e3 & 1.65e5 & 22.87$\pm$0.05 &  &
    3524.5 & 3.05e3 & 1.56e5 & 23.96$\pm$0.03 \\
    3823$^b$ & 1.69e3 & 4.12e4 & 23.37$\pm$0.15 \\
    \hline
    \multicolumn{9}{l}{Notes: $^a$ The FeI line designated 3746 includes lines at 3743.4, 3745.6, 3745.9, 3748.3, and 3749.5\AA } \\
    \multicolumn{9}{l}{\hspace{0.5in}$^b$ The FeI line designated 3823 includes lines at 3815.8, 3820.4, 3824.4, 3825.9, and 3827.8\AA } \\
    \multicolumn{9}{l}{\hspace{0.6in}$v$=0.3 km/s used to set all scale lengths} \
    
 \end{tabular}

\end{table*}
Inspection of Table~\ref{feniscales} shows that the parent scale lengths are short relative to the scale lengths for the molecular species. Indeed, the parent scale lengths are all similar, at $\sim3\times10^3$\,km/s.  This short parent scale length is comparable to the size of a pixel. We do not have any information below this size.  It would be possible for the parent scale length to be even shorter, possibly just above the surface of the nucleus.

We assumed an outflow velocity of the gas of 0.3\,km/s to fit the data and derive the daughter scale lengths.  We could not find daughter scale lengths that produced a model curve of a proper shape to fit the data points farther from the optocenter using the nominal Solar System comet outflow velocities instead of these seemingly slow outflow velocities.  The velocity is the same as was found with the radio observations above.  We were only able to constrain the daughter scale lengths because of the relatively large FOV of the VIRUS IFU.  However, inspection of Table~\ref{feniscales} shows that there are discrepancies in the daughter scale lengths  derived for different examples of each species.  This is the result of even our large FOV being smaller than the needed daughter scale lengths. UVES and KCWI both had too small FOVs to determine these scale lengths  (The FOV of KCWI was $8" \times 20"$ and for UVES was $1.8" \times 7.5"$,
whereas our FOV was $50" \times 50"$ arcsec.)  

\citet{hutsetal2026} observed 3I at high spectral resolving power as it approached perihelion.  They first  observed NiI at 3.78\,au inbound.  However, they did not observe any FeI until the comet reached 2.64\,au.  \citet{Rahatgaonkaretal2025} and \citet{2025arXiv251011779H} also detected NiI at larger heliocentric distances before FeI was detected pre-perihelion.  

Averaging our FeI and NiI observations for all lines included in our Table~\ref{feniscales}, we find a value for log(Q(Ni)/Q(Fe)=0.19, similar to the values reported in \citet{HutetalFeNiorigin2026} for heliocentric distances comparable to our observations.

\section{Conclusions}
The excellent spatial resolution of VIRUS, coupled with a FOV of $50\times50$\,arcsecs and a geocentric distance of 1.83\,au, allowed us to probe the coma of 3I in great detail. 
The accidental offset of the optocenter of the comet to the edge of the FOV meant that we probed to a large distance ($>$85,000km) from the optocenter.  This added new information to the extensive literature on 3I, even with a single epoch of observations.  With the spatial information that we were able to obtain, we could trace the outflow of the gas, including molecular species and Fe and Ni.  Assuming the molecular lifetimes are normal in the vicinity of the Sun, we concluded that the gas must be flowing outwards from the optocenter much more slowly than in the comae of typical comets, in agreement with radio telescope measurements of parent species' outflow velocities. 

The presence of very strong NiI and FeI lines in spectra of 3I is an important difference from Solar System comets.  3I was exceptional even 
compared to the other interstellar objects, 1I and 2I, due to its exceptionally high NiI and FeI \citep{hutsetal2026}.
Wandering around in the interstellar medium should result in the body having high material strength, with FeI and NiI metal on its surface \citep{TrigRod2025}. 
However, the surface temperatures at the heliocentric distances that 3I was observed are too low to vaporize Fe or Ni in metallic form \citep{manetalFeNi2021}.  The very short parent scale lengths that we measure must indicate that Fe and Ni are coming directly off the nucleus in their metallic form
or must have a very short-lived parent.  This was 
also the conclusion about the origin of the FeI and NiI observed at high spectral resolving power 
\citep{manetalFeNi2021, hutsetal2026}.
Indeed, \citet{2025arXiv251119112T} suggested that 3I has a composition that is a close match to a carbonaceous chondrite and might be the result of its elevated metallic abundance and abundant water ice.

If comets are leftovers from the formation of most planetary systems, and a high fraction are ejected from their formation nebula, there should be many such objects wandering around the
Galaxy that can potentially cross through our Solar System. Our Sun, like most planetary systems, probably formed in a cluster of infant stellar systems. The leftover debris of all those stars can, therefore,  be influenced by nearby passing stellar systems. As the young stars orbit around the center of the Galaxy, the debris they have ejected will sometimes continue wandering in space, devoid of a star to which they are attached.  Thus, unlike a comet in our Solar System that can be influenced by the gravity and radiation of our Sun, these other interstellar comet-like objects may be going into and out of insolation from stars they pass.

The low outflow velocities, and therefore the short scale lengths, that we found for the typical cometary gas molecules also indicate something is different  for 3I compared with normal comets. Once the ices have sublimed to the gas form, they should respond to sunlight in the same manner as for our Solar System comets if the parents are identical.  However, with a very different parent star, it is possible that there are differences in the composition of the ices.  As noted above, 3I is rich in CO$_2$.  \citet{biveretal2026} suggested that a heavy molecular gas, such as CO$_2$-rich gas, is what is slowing the outflow velocity, since CO$_2$ is much heavier than the predominant gas in the coma of Solar System comets.

As new telescopes come on line (e.g. the Rubin telescope 
\citep{RubinScience}) with large surveys, we should expect that the number of interstellar objects detected passing through our Solar System will increase.  Observations of these objects will yield important comparative information about how planetary systems form and potentially lose or gain material.  Observations of 3I demonstrate the need for a wide variety of types of observations to try to understand similarities or differences with our solar nebula.  Ultimately, not being sure where or when these objects formed will be our biggest impediment to using interstellar objects to constrain the formation of our Solar System.


\begin{acknowledgements}
We thank Sergey Rostopchin, Zachary Vanderbosch, Jose Fernandez, and Justen Pautzke, all members of the staff of the Hobby-Eberly Telescope, for their assistance with specifying and applying the necessary files to obtain VIRUS data on 3I.

VIRUS is a joint project of the University of Texas at Austin,
Leibniz-Institut f\"ur Astrophysik Potsdam (AIP), Texas A\&M University
(TAMU), Max-Planck-Institut f\"ur
Extraterrestrische Physik (MPE),
Ludwig-Maximilians-Universit\"at Muenchen, Pennsylvania State
University, Institut f\"ur Astrophysik G\"ottingen, University of Oxford,
and the Max-Planck-Institut f\"ur Astrophysik (MPA). In addition to
Institutional support, VIRUS was partially funded by the National
Science Foundation, the State of Texas, and generous support from
private individuals and foundations.
\end{acknowledgements}




%
\facility{HET}

\bibliographystyle{aasjournal}
\bibliography{HET_ATLAS_3I.bib}

\end{document}